\documentclass[10pt,final,conference,a4paper]{IEEEtran}
\usepackage{tikz}
\usepackage{amsmath,amsfonts,amssymb,bm,mathrsfs}

\usetikzlibrary{shapes,arrows}
\definecolor{olive}{rgb}{0.3, 0.4, .1}
\definecolor{rose}{rgb}{1., 0.75, .75}
\definecolor{lightyellow}{rgb}{1., 1, .5}
\definecolor{lightblue}{rgb}{.95, 95, 1}
\definecolor{lightcyan}{rgb}{.5, 1, 1}
\definecolor{sectionblue}{rgb}{.20 .20, .702}
\definecolor{midblue}{rgb}{.65 .75, 1}
\definecolor{darkblue}{rgb}{0.1, 0.1, 0.4}
\definecolor{brown}{rgb}{.4, .2, 0}

\begin{document}
\title{Least square estimation of phase, frequency and PDEV}
\author{Magnus Danielson, Fran\c{c}ois Vernotte, Enrico Rubiola}
\author{\IEEEauthorblockN{Magnus Danielson}
\IEEEauthorblockA{R\&D System Design\\
Net Insight AB\\
Stockholm, Sweden\\
Email: magda@netinsight.net}
\and
\IEEEauthorblockN{Fran\c{c}ois Vernotte}
\IEEEauthorblockA{Observatory THETA/UTINAM,\\UBFC/UFC and CNRS\\
Besan\c{c}on, France\\
Email: francois.vernotte@obs-besancon.fr}
\and
\IEEEauthorblockN{Enrico Rubiola}
\IEEEauthorblockA{CNRS FEMTO-ST Institute,\\Dept Time and Frequency\\
Besan\c{c}on, France\\
Email: rubiola@femto-st.fr}}
\maketitle

\begin{abstract}
The $\Omega$ pre\-processing was introduced to improve phase noise rejection by using a least square algorithm. The associated variance is the PVAR which is more efficient than MVAR to separate the different noise types. However, unlike AVAR and MVAR, the decimation of PVAR estimates for multi-$\tau$ analysis is not possible if each counter measurement is a single scalar. This paper gives a decimation rule based on two scalars, the processing blocks, for each measurement. For the $\Omega$ pre\-processing, this implies the definition of an output standard as well as hardware requirements for performing high-speed computations of the blocks.
\end{abstract}

\section{Introduction\label{sec:new_intro}}
\subsection{Background}
Allan variance (AVAR) \cite{allan1966} and the log-log plot of Allan deviation (ADEV) is the
standard tool since its introduction in 1966, and still relied on the in the time
and frequency community. With the introduction of Parabolic Variance (PVAR)
\cite{rubiola2015} an improved tool was presented, potentially more suited to
practical problems for fast processes.

It's benefits lies in that it can separate white PM from flicker PM and it
rejects white PM of the counter input stage with $1/\tau^3$ in the variance.
The white PM rejection is relevant as it integrates over $0.5-5 GHz$ on some
counters, thus having a high noise bandwidth.

PVAR combines the advantage of both methods, but comes at the cost of
processing power needed, which could be addressed using FPGA technology for
high speed sample gathering and decimation prior to further processing in
software.

The major reservation to the use of PVAR has been that no decimation
rule was known in the T\&F community, thus the full series of phase-time data
had to be stored and processed for each value of $\tau$ in the PDEV log-log
plot. The evaluation of frequency over 1 day takes storing $10^{11}$ phase-time
data sampled at $1 \mu s$ interval.

This article presents the decimation rule needed, allowing for significant
reduction in memory and processing needs, thus providing means for making the
PVAR processing practical and useful.
For each value of $\tau$ we only need a short series data (two scalars) to be
stored. The maximum record length is limited by the confidence level desired,
$10^3$ being probably large enough for virtually all practical purposes.
The algorithm is surprisingly simple because the least square estimator relies
on linear operators only. This also allows the decimator rule to be applied
recursively for further reductions as needed in software for longer $\tau$
processing.

\subsection{No pre\-processing}
Traditional time-interval and frequency counters provided no pre\-processing,
even if average by $N$ was possible to select. A time-interval counter
produces a sequence of phase difference samples, while the frequency read-out
produces the difference between these phase differences divided by the time
between them. The improvement of counters lay in the improved resolution of
the single-shot resolution and the reduction of trigger noise in each such
measurement, thus reducing both the systematic and random noise processes in
the measurements. Another major development was the ability to make continuous
measurement, where samples will be collected without a dead-time in between the
last sample of the previous measure and the first of the next measure, but
where this is the same sample.

\subsection{$\Lambda$ pre\-processing and $\Lambda$-counters}
In order to meet the challenges of white noise limitations to measure while
measuring optical beat frequencies, Snyder \cite{snyder1981} introduced a
method to pre-filter samples in order to improve the noise rejection,
achieving a deviation having the slope of $1/\tau^{1.5}$ over the traditional
$1/\tau$ slope of white noise reduction, where $\tau$ is the time
between phase observations, this providing a much improved filtering for the
same $N$ samples being averaged. Snyder also presents a hardware accumulation
that allows for such improved frequency observations so that a high rate of
observations can be accumulated in high rate by hardware, and only a post\-processing need to be achieved in software. This have since been introduced
into commercial counters as means to increase the frequency reading precision
compared to the update rate.

\subsection{Effect on variance estimation}
The use of different frequency estimator pre\-processing support in counters
has shown to have an impact on the estimation of Allan variance (AVAR), as
shown by Rubiola \cite{rubiola2005rsi}. Applying the $\Lambda$ pre\-processing
to Allan variance processing produces a variance known as a Modified Allan
 variance (MVAR) \cite{allan1981} as presented by Allan.
However, in order to extend the pre\-processing properties to get the proper
MVAR, the data must be decimated properly. This requires us to distinguish
the different type of pre\-processing. Rubiola introduced the term $\Pi$-counter
for the classical not pre\-processed response, as it present an evenly weighting of
the frequency, and the weight function graphically looks similar to the
$\Pi$ sign. Similarly, the weight function on frequency for the pre\-processing method of Snyder is referred to as a $\Lambda$-counter, as it is produced by
the $\Lambda$ pre\-processing.
\begin{table*}[tbp]\centering
\caption{Allan, Modified Allan and Parabolic variance of the different noise
forms.\label{tab:noisetable}}
\begin{center}
\begin{tabular}{|l|c|c|c|c|}      \hline
\multicolumn{2}{|c|}{Variance Type} & AVAR & MVAR & PVAR \\ \hline
\multicolumn{2}{|c|}{Preprocessing type} & $\Pi$ & $\Lambda$ & $\Omega$ \\ \hline
Noisetype & $S_y(f)$ & & & \\ \hline
& & & & \\
White Phase & $h_2f^2$ & $\frac{3f_H}{4\pi^2\tau^2}h_2$ & $\frac{3}{8\pi^2\tau^3}h_2$ & $\frac{3}{2\pi^2\tau^3}h_2$ \\
& & & & \\ \hline
& & & & \\
Flicker Phase & $h_1f$ & $\frac{3[\gamma+\ln(2\pi f_H\tau)]-\ln2}{4\pi^2\tau^2}h_1$ & $\frac{24\ln2-9\ln3}{8\pi^2\tau^2}h_1$ & $\frac{3[\ln(16)-1]}{2\pi^2\tau^2}h_1$ \\
& & & & \\ \hline
& & & & \\
White Frequency & $h_0$ & $\frac{1}{2\tau}h_0$ & $\frac{1}{4\tau}h_0$ & $\frac{3}{5\tau}h_0$ \\
& & & & \\ \hline
& & & & \\
Flicker Frequency & $h_{-1}f^{-1}$ & $2\ln(2) h_{-1}$ & $\frac{27\ln(3)-32\ln(2)}{8}h_{-1}$ & $\frac{2[7-\ln(16)]}{5}h_{-1}$ \\
& & & & \\ \hline
& & & & \\
Random Walk Frequency & $h_{-2}f^{-2}$ & $\frac{2\pi^2\tau^2}{3}h_{-2}$ & $\frac{11\pi^2\tau}{20}h_{-2}$ & $\frac{26\pi^2\tau}{35}h_{-2}$ \\
& & & & \\ \hline
\multicolumn{5}{|l|}{$\gamma=0.577215$, the Euler-Mascheroni constant} \\ \hline
\end{tabular}
\end{center}
\end{table*}

Table \ref{tab:noisetable} give the formulas for the variance of different pre\-processing types and hence variance types. Notice how White Phase Modulation
has a different slope for MVAR and PVAR compared to AVAR, this illustrate the
improved white noise rejection of these variances compared to no pre\-processing.

\subsection{Pre\-processing filters}
For $\Pi$-counters, we can always produce AVAR and MVAR. For $\Lambda$-counters
(a counter in it's $\Lambda$ pre-filtering mode), we can only produce proper
MVAR results with proper decimation of data. Just using the frequency
estimates in Allan Variance will not provide proper results, but biased
results, where the bias decreases for longer $\tau$ as the fixed bandwidth of
the counters pre\-processing wears off as the Allan Variance itself has a
filtering effect. The filtering thus represents the effect of a low-pass
filter, lowering the system bandwidth and hence the systems sensitivity to
white noise. Proper decimation requires that the decimation routines process
data such that the filtering continues to reduce the bandwidth of this filter
as data is combined for longer observation periods, and thus maintain the
benefit of such processing.

\subsection{$\Omega$ pre\-processing and $\Omega$-counters\label{sec:Omega_preprocess}}
This paper concerns itself with the decimation of data in a third type of
counter known as the $\Omega$-counter, thus a counter having a frequency
weight function looking similar to the $\Omega$ sign. This is a parabolic
curve which is the result of using a least-square estimation of the frequency
slope out of the phase data. This processing produces a new type of variance
known as the Parabolic Variance (PVAR) and has even better properties with
regard to suppressing the white noise. However, the \cite{rubiola2015} gave no guidance to
an algorithm of decimation, or how the hardware accumulation should be done
such that performance benefits can be achieved for any multiple of such block
length. 

This paper develops a discrete time estimators and then decimation
methods such that high speed accumulation can be used together with post\-processing to achieve memory and computational efficient processing for
multi-$\tau$ PDEV log-log plots, this without altering the properties of PVAR as given in \cite{rubiola2015} and Table \ref{tab:noisetable}. After reminding the main features of the $\Omega$-counters and of the PVAR ({\S} \ref{sec:Omega_and_PVAR}), the basics of decimation 
will be presented in section {\S} \ref{sec:decimAM}. Unfortunately, it turns out that the decimation is not a trivial problem with $\Omega$-counters and then with PVAR. But a simple solution will be given in {\S} \ref{sec:soldec} after having recalled the basics of least squares ({\S} \ref{sec:ls}). Finally, recommendation will be given for choosing a standard for the output format of the $\Omega$-counters in {\S} \ref{sec:standard}.

\section{PVAR and $\Omega$-counters\label{sec:Omega_and_PVAR}}
The concept of $\Omega$-counter was formulated by Rubiola \cite{rubiola2015}, based on Johansson \cite{johansson2005}, to achieve the optimal rejection of white phase noise for short term frequency measurement by using an estimator based on the least squares. Such methods was presented by Barnes \cite{barnes1983},
for the purpose of drift estimation under presence of white noise, but
Johansson \cite{johansson2005} makes the first connection between least square
methods and AVAR, but without providing the effect on various noise-types and
related bias functions.

The principle of this frequency estimation is to calculate the least squares slope over a phase sequence $\left\{x_k\right\}$ obtained at instants $t_k=k\tau_0$ with $k\in\left\{0,\ldots,N-1\right\}$ where $\tau_0$ is the sampling step and $\tau=(N-1)\tau_0$ the total length of the sequence. It is well known that the least squares provide the best slope estimate in the presence of white noise (i.e. white PM noise) \cite{barnes1983}. Such an estimate, that we denote $\hat{\mathbf{y}}^\Omega$, is obtained by a weighting average of the phase data:
\begin{eqnarray}
\hat{\mathbf{y}}^\Omega&=&\int_{0}^{\tau} \tilde{w}_c(t-\frac{\tau}{2}) x(t)\,dt\label{eq:y_om_cont}\\
\hat{\mathbf{y}}^\Omega&\simeq&\frac{1}{\tau_0}\sum_{k=0}^{N-1} \tilde{w}_c(t_k-t_{N/2}) x_k\label{eq:y_om}
\end{eqnarray}
where the phase weight function $\tilde{w}_c(t)$ is defined as:
\begin{equation}
\left\{
\begin{array}{lcl}
\tilde{w}_c(t)=\displaystyle\frac{12}{\tau^3}t & \quad & \text{if} \quad t \in \left[-\tau/2,+\tau/2\right]\\
\tilde{w}_c(t)=0 & \quad & \text{elsewhere.}
\end{array}
\right.\label{eq:wcx}
\end{equation}
The estimator for sample data (\ref{eq:y_om}) is here given as an approximate,
but a bias free variant will be presented in the paper.
It has been demonstrated that, in the presence of white PM, the variance of this frequency estimate is lower  by  a  factor  of $\frac{3}{4}$ than the variance of the corresponding $\Lambda$-counter estimate. Moreover, since the least squares are optimal for white noise, the variance of the $\Omega$-counter estimate is minimal. It is then an efficient estimator \cite{everitt1998}.

This estimator may be also computed from frequency deviation samples $\left\{\bar{y}_k\right\}$ defined as $\bar{y}_k=\frac{x_{k+1}-x_k}{\tau_0}$:
\begin{eqnarray}
\hat{\mathbf{y}}^\Omega&=&\int_{0}^{\tau} w_c(t-\frac{\tau}{2}) \bar{y}(t)\,dt\\
\hat{\mathbf{y}}^\Omega&\simeq&\sum_{k=0}^{N-1} w_c(t_k-t_{N/2}) \bar{y}_k\label{eq:yy_om}
\end{eqnarray}
where the frequency weight function $w_c(t)$ is defined as:
\begin{equation}
\left\{
\begin{array}{ll}
w_c(t)=\displaystyle\frac{3}{2\tau}\left[1-\frac{4t^2}{\tau^2}\right] & \text{if} \quad t \in \left[-\tau/2,+\tau/2\right]\\
w_c(t)=0 & \text{elsewhere.}
\end{array}
\right.\label{eq:wcy}
\end{equation}
The estimator for sample data (\ref{eq:yy_om}) is again given as an approximate,
but a bias free variant will be presented in the paper.

\begin{figure}
A \includegraphics[height=2.7cm]{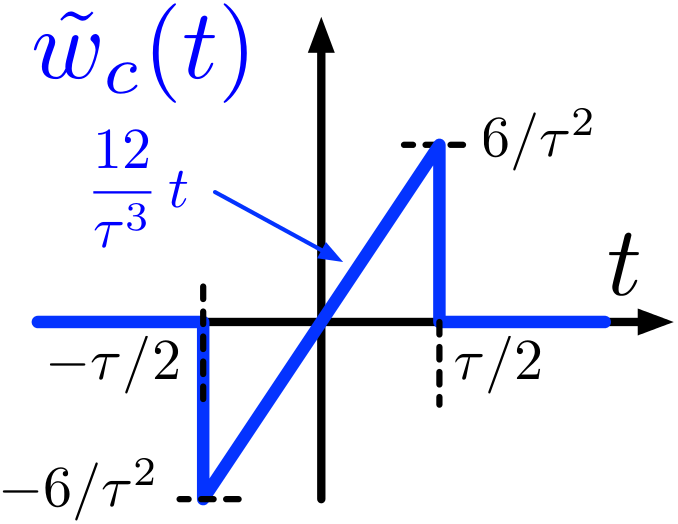}
\hfill B \includegraphics[height=2.7cm]{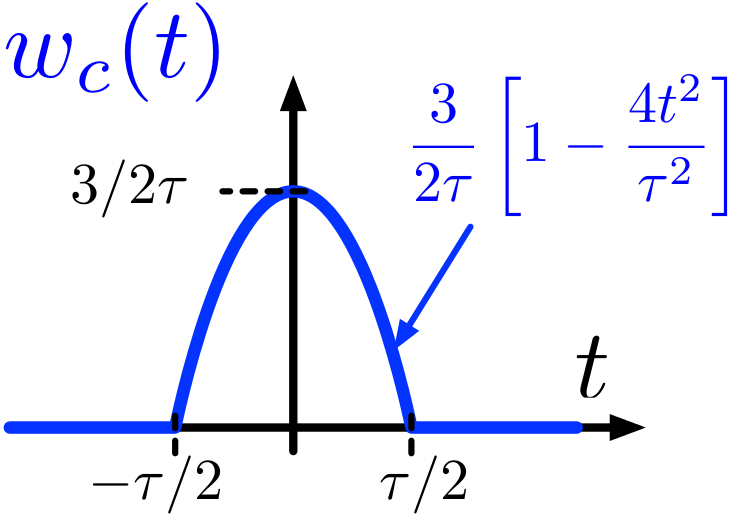}
\caption{weight functions of the $\Omega$-counter computed from phase data (A, left) or from frequency deviations (B, right).\label{fig:wc_xny}}
\end{figure}

The $\Omega$-counter weight functions for phase data as well as for frequency deviations are plotted in Figure \ref{fig:wc_xny}. The shape of $w_c(t)$ (see Figure \ref{fig:wc_xny}-B) explains the choice of the Greek letter $\Omega$ to name this counter \cite{rubiola2005rsi,rubiola2015}.

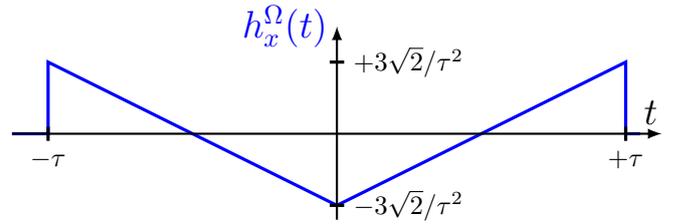
\begin{figure}
\begin{center}
\begin{tikzpicture}
\begin{scope}[scale=0.95]
\draw[very thick,blue] (0.5,0)--(1,0)--(1,1)--(5,-1)--(9,1)--(9,0)--(9.2,0);

\draw[thick,->,-latex] (0.5,0)--(9.5,0);
\draw (9.35,0) node[above]{\Large$t$};
\draw[thick,->,-latex] (5,-1.2)--(5,1.5);
\draw[blue] (5,1.5) node[left]{{\Large $h_x^\Omega(t)$}};
\draw[very thick] (1,-0.1)--(1,+.1);
\draw (1,-.1) node[below]{$-\tau$};
\draw[very thick] (9,-0.1)--(9,+.1);
\draw (9,-.1) node[below]{$+\tau$};
\draw[very thick] (4.9,-1)--(5.1,-1);
\draw (5.1,-1) node[right]{$-3\sqrt{2}/\tau^2$};
\draw[very thick] (4.9,1)--(5.1,1);
\draw (5.1,1) node[right]{$+3\sqrt{2}/\tau^2$};
\end{scope}
\end{tikzpicture}
\end{center}
\caption{weight function associated to PVAR for phase data.\label{fig:wPVARx}}
\end{figure}

For each type of counter, a specific statistical estimator has been defined for stability analysis: AVAR for $\Pi$-counters \cite{allan1966}, the traditional time-interval or frequency counter, and MVAR for $\Lambda$-counters \cite{allan1981}, as inspired by the work of Snyder \cite{snyder1981}. In the same way, the Parabolic variance (PVAR) was defined to handle $\Omega$-counter measurements \cite{vernotte2015, benkler2015}. The general relationship defining a $X$-variance ($X$ being A, M or P) from a $\chi$-counter ($\chi$ being $\Pi$, $\Lambda$ or $\Omega$) is:
$$
X\text{VAR}(\tau)=\frac{1}{2}\left<\left(\hat{\mathbf{y}}^{\chi}_{2}-\hat{\mathbf{y}}^{\chi}_{1}\right)^2\right>
$$
where $\hat{\mathbf{y}}^{\chi}_{1}$ is the frequency estimate given by a $\chi$-counter at instant $t_1$ and $\left<\cdot\right>$ stands for an ensemble average over all available frequency estimates. In this connection, PVAR is then defined as PVAR$(\tau)=\frac{1}{2}\left<\left(\hat{\mathbf{y}}^{\Omega}_{2}-\hat{\mathbf{y}}^{\Omega}_{1}\right)^2\right>$ \cite{vernotte2015}. The weight function associated to PVAR for phase data is plotted in Figure \ref{fig:wPVARx}.

PVAR, like MVAR, is intended to deal with short term analysis (and then white and flicker PM noises) whereas AVAR is preferred for the measurement of long term stability and timekeeping. The main advantage of PVAR regarding MVAR relies on the larger EDF of its estimates, and in turn the smaller confidence interval. The best of PVAR is its power to detect and identify weak noise processes with the shortest data record.  PVAR is superior to MVAR in all cases, and also superior to AVAR for all short-term and medium-term processes, up to flicker FM included.  AVAR is just a little better with random walk and drift. Therefore, PVAR should be an improved replacement for MVAR in all cases, provided the computing overhead can be accepted.  

Thus, the only drawback of PVAR lies in the difficulty to find its decimation algorithm as stated in {\S} \ref{sec:Omega_preprocess}. In order to solve this problem, let us remind the basics of decimation.

\section{Decimation with AVAR and MVAR\label{sec:decimAM}}
	\subsection{AVAR}
The frequency estimate given by a $\Pi$-counter over the time interval $\tau=N\tau_0$ beginning at the instant $t_k$ is 
$$
\hat{\mathbf{y}}^{\Pi}_{k,\tau}=\frac{x_{k+N}-x_k}{\tau}.
$$
Therefore, AVAR$(\tau)$ is obtained from
\begin{eqnarray}
\text{AVAR}(\tau)&=& \frac{1}{2}\left<\left(\hat{\mathbf{y}}^{\Pi}_{k+N}-\hat{\mathbf{y}}^{\Pi}_{k}\right)^2\right>\nonumber\\
&=&\frac{1}{2\tau^2}\left<\left(x_{k+2N}-2x_{k+N}+x_k\right)^2\right>\nonumber
\end{eqnarray}
where the ensemble average is performed over all $k$ values (overlapped AVAR). 

The passage from $\tau$ to $2\tau$ is quite obvious: 
$$
\hat{\mathbf{y}}^{\Pi}_{k,2\tau}=\frac{x_{k+2N}-x_k}{2\tau}=\frac{\hat{\mathbf{y}}^{\Pi}_{k,\tau}+\hat{\mathbf{y}}^{\Pi}_{k+N,\tau}}{2}
$$
and then
$$
\text{AVAR}(2\tau)= \frac{1}{2}\left<\left(\frac{\hat{\mathbf{y}}^{\Pi}_{k+3N}+\hat{\mathbf{y}}^{\Pi}_{k+4N}-\hat{\mathbf{y}}^{\Pi}_{k}-\hat{\mathbf{y}}^{\Pi}_{k+N}}{2}\right)^2\right>.
$$
This decimation rule can be easily extended to any multiple of $\tau$. Therefore, the knowledge of a sequence of contiguous $\left\{ \hat{\mathbf{y}}^{\Pi}_{k,\tau}\right\}$ with $k\in\left\{0,\cdots,N-1\right\}$ allows us to compute AVAR for any multiple of $\tau$ up to $N\tau/2$.

	\subsection{MVAR}
The frequency estimate given by a $\Lambda$-counter over the time interval $\tau=N\tau_0$ beginning at the instant $t_k$ is 
\begin{equation}
\hat{\mathbf{y}}^{\Lambda}_{k,\tau}=\frac{1}{N\tau}\sum_{j=0}^{N-1}\left(x_{k+j+N}-x_{k+j}\right)\label{eq:y_lambda}.
\end{equation}
Thus
\begin{eqnarray}
\text{MVAR}(\tau)&=& \frac{1}{2}\left<\left(\hat{\mathbf{y}}^{\Lambda}_{k+N}-\hat{\mathbf{y}}^{\Lambda}_{k}\right)^2\right>\label{eq:MVAR}\\
&=&\frac{1}{2N^2\tau^2}\left<\left[\sum_{j=0}^{N-1}\left(x_{k+j+2N}\right.\right.\right.\nonumber\\
&&\left.\left.\left. -2x_{k+j+N}+x_{k+j}\right)\vphantom{\sum_{j=0}^{N-1}}\right]^2\right>.\nonumber
\end{eqnarray}
The passage from $\tau$ to $2\tau$ is given by:
\begin{eqnarray}
\hat{\mathbf{y}}^{\Lambda}_{k,2\tau}&=&\frac{1}{2N\tau}\sum_{j=0}^{2N-1}\left(x_{k+j+2N}-x_{k+j}\right)\nonumber\\
&=&\hat{\mathbf{y}}^{\Lambda}_{k,\tau}+2\hat{\mathbf{y}}^{\Lambda}_{k+N,\tau}+\hat{\mathbf{y}}^{\Lambda}_{k+2N,\tau}.\nonumber
\end{eqnarray}
More generally, it can be demonstrated that the general decimation rule is:
\begin{eqnarray}
\hat{\mathbf{y}}^{\Lambda}_{k,n\tau}&=&\sum_{i=0}^{n-2}(i+1)\left(\hat{\mathbf{y}}^{\Lambda}_{k+iN,\tau}+\hat{\mathbf{y}}^{\Lambda}_{k+(2n-2-i)N,\tau}\right)\nonumber\\
&&+n\hat{\mathbf{y}}^{\Lambda}_{k+(n-1)N,\tau}.\nonumber
\end{eqnarray}
Here also, the knowledge of a sequence of contiguous $\left\{ \hat{\mathbf{y}}^{\Lambda}_{k,\tau}\right\}$ with $k\in\left\{0,\cdots,N-1\right\}$ allows us to compute MVAR for any multiple of $\tau$ up to $N\tau/3$.

	\subsection{PVAR}
The frequency estimate given by an $\Omega$-counter over the time interval $\tau=N\tau_0$ beginning at the instant $t_k$ is given by (\ref{eq:y_om}) and (\ref{eq:wcx}). But in this case, no decimation rule may be found for passing from $\tau$ to a multiple $n\tau$ for any $n\in\mathbb{N}$. In order to get a decimation rule, we will demonstrate that an $\Omega$-counter must provide 2 scalars. Let us go back to the basics of the least squares to better understand this issue.

\section{Least-square frequency estimation\label{sec:ls}}
\subsection{Linear system}
The least square system producing the output vector $\mathbf{x}$ of
phase samples from the system state vector $\mathbf{c}$ using the system matrix
$\mathbf{A}$ and assuming the error contribution of $\mathbf{d}$ as defined in
$\mathbf{x} = \mathbf{A}\mathbf{c} + \mathbf{d}$
having the least square estimation as given by
\begin{equation}
\hat{\mathbf{c}} = (\mathbf{A}^T\mathbf{A})^{-1}\mathbf{A}^T\mathbf{x}\label{eq:lsestimator}
\end{equation}
For this system, a linear model of phase and frequency state is defined
\begin{equation}
\hat{\mathbf{c}} = \left(
\begin{array}{c}
\hat{x}\\\hat{y}
\end{array}
\right)\label{eq:eststate}
\end{equation}
A block of phase samples, taken with $\tau_0$ time in-between them,
building the series $x_n$ where $n$ is in the range $\left\{0,\ldots, N-1\right\}$ where by
convention $N$ is the number of phase samples. In the system model, each sample
$n$ has an associated observation time $t_n=\tau_0n$.  The matrix $\mathbf{A}$ and the vector $\mathbf{x}$ then becomes
\begin{eqnarray}
\mathbf{A}&=&\left(\begin{array}{cc}1&t_0\\
\vdots&\vdots\\
1&t_n\\
\vdots&\vdots\\
1&t_{N-1}\end{array}\right) = \left(\begin{array}{cc}1&0\\
\vdots&\vdots\\
1&\tau_0n\\
\vdots&\vdots\\
1&\tau_0(N-1)\\
\end{array}\right)\label{eq:Amatrix}\\
\mathbf{x}&=&\left(\begin{array}{c}x_0\\
\vdots\\
x_n\\
\vdots\\
x_{N-1}\end{array}\right)\label{eq:xmatrix}
\end{eqnarray}
\subsection{Closed form solution}
Inserting (\ref{eq:Amatrix}) and (\ref{eq:xmatrix}) into (\ref{eq:lsestimator})
results in
\begin{eqnarray}
\mathbf{\hat{c}}&=&
\left[
\left(\begin{array}{ccc}
\ldots &1&\ldots\\
\ldots &\tau_0n&\ldots
\end{array}\right)
\left(\begin{array}{cc}
\vdots & \vdots\\
1&\tau_0n\\
\vdots & \vdots\\
\end{array}\right)
\right]^{-1}\nonumber\\
&&\times
\left(\begin{array}{ccc}
\ldots &1&\ldots\\
\ldots &\tau_0n&\ldots
\end{array}\right)
\left(\begin{array}{c}
\vdots\\
x_n\\
\vdots\\
\end{array}\right)
\end{eqnarray}
simplifies into
\begin{equation}
\mathbf{\hat{c}}=
\left(\begin{array}{cc}
\displaystyle \sum_{n=0}^{N-1}1 & \displaystyle \tau_0\sum_{n=0}^{N-1}n\\
\displaystyle \tau_0\sum_{n=0}^{N-1}n & \displaystyle \tau_0^2\sum_{n=0}^{N-1}n^2
\end{array}\right)^{-1}
\left(\begin{array}{c}
\displaystyle \sum_{n=0}^{N-1}x_n\\
\displaystyle \tau_0\sum_{n=0}^{N-1}nx_n\\
\end{array}\right)
\label{eq:syssolve}\end{equation}
replacing the sums $C$ and $D$
\begin{eqnarray}
C&=&\sum_{n=0}^{N-1}x_n\label{eq:lsCsum}\\
D&=&\sum_{n=0}^{N-1}nx_n\label{eq:lsDsum}
\end{eqnarray}
becoming
\begin{equation}
\mathbf{\hat{c}}=
\left(\begin{array}{cc}N&\tau_0\frac{N(N-1)}{2}\\\tau_0\frac{N(N-1)}{2}&\tau_0^2\frac{N(N-1)(2N-1)}{6}\end{array}\right)^{-1}
\left(\begin{array}{cc}C\\\tau_0D\end{array}\right)\label{eq:lssolution}
\end{equation}
inverse can be solved as
\begin{eqnarray}
\left(\begin{array}{cc}N&\tau_0\frac{N(N-1)}{2}\\\tau_0\frac{N(N-1)}{2}&\tau_0^2\frac{N(N-1)(2N-1)}{6}\end{array}\right)^{-1}\nonumber\\
={\frac{12}{\tau_0^2N(N-1)(N+1)}}\nonumber\\
\times \left(\begin{array}{cc}\tau_0^2\frac{(N-1)(2N-1)}{6}&-\tau_0\frac{N-1}{2}\\-\tau_0\frac{N-1}{2}&1\end{array}\right)
\label{eq:sysarray}\end{eqnarray}
insertion of (\ref{eq:eststate}) and (\ref{eq:sysarray}) into (\ref{eq:lssolution}) resulting in the estimators
\begin{eqnarray}
\hat{x}=&\displaystyle \frac{6}{N(N+1)}\left(\frac{(2N-1)}{3}C-D\right)\label{eq:lsphase}\\
\hat{y}=&\displaystyle \frac{12}{\tau_0N(N-1)(N+1)}\left(-\frac{N-1}{2}C+D\right)\label{eq:lsfreq}
\end{eqnarray}
these estimators have been verified to be bias free from static phase and
static frequency, as expected from theory. Using these estimator formulas
the phase and frequency can estimated of any block of $N$ samples for which
the $C$ and $D$ sums have been calculated.

\subsection{LS weight function}
From (\ref{eq:lsfreq}), we can write $w_{y,x}(n)$, the weight function of the unbiased LS frequency estimator on $\left\{x_n\right\}$ data:
\begin{eqnarray}
\hat{y}&=&\sum_{n=0}^{N-1}\frac{12}{\tau_0N(N-1)(N+1)}\left(-\frac{N-1}{2}+n\right)x_n\nonumber\\
&=&\sum_{n=0}^{N-1}w_{y,x}(n) x_n
\end{eqnarray}
with 
$$
w_{y,x}(n) = \frac{12}{\tau_0N(N-1)(N+1)}\left(-\frac{N-1}{2}+n\right)
$$

Similarly, the weight functions of the unbiased LS frequency estimator on frequency samples become
\begin{eqnarray}
\hat{y}&=&\sum_{m=0}^{N-2}w_{y,y}(m)y_m\\
w_{y,y}(m)&=&\frac{6(N-1-m)(m+1)}{\tau_0N(N-1)(N+1)}
\end{eqnarray}
These weight functions are different from $\tilde{w}_c(t)$ and $w_c(t)$ introduced in (\ref{eq:wcx}) and (\ref{eq:wcy}) because the later are centered and calculated in a continuous case (see \cite{rubiola2015}). It can be easily verified that:
$$
\lim_{N\gg 1} w_{y,x}(n)=\frac{1}{\tau_0}\tilde{w}_c\left(t_n\right) \quad \text{and} \quad \lim_{N\gg 1} w_{y,y}(n)=w_c\left(t_n\right)
$$
with $t_n=(n-(N-1)/2)\tau_0$ for centering the dates. The continuous time
definition does not compare easilly to those of discrete time, where the
discrete time have the terms $(N-1)(N+1)$ rather than $N^2$ in order to be
bias-free.
Therefore, for a lower number of samples $N$, $w_{y,x}(n)$ and $w_{y,y}(n)$ should be used to avoid estimator biases (see {\S} \ref{sec:bias}). 

\subsection{PVAR calculation}
The PVAR estimator calculation is defined from the equations
\begin{eqnarray}
\hat{\sigma}_P^2(\tau)&=&\frac{1}{M}\sum_{i=1}^M(\alpha_i)^2\label{pvaralphasum}\\
\alpha_i&=&\frac{1}{\sqrt2}\left(\hat{\mathbf{y}}_i^\Omega-\hat{\mathbf{y}}_{i+1}^\Omega\right)\label{eq:pvaralpha}
\end{eqnarray}
inserting (\ref{eq:lsfreq}) and (\ref{eq:pvaralpha}) into (\ref{pvaralphasum}) produces
\begin{eqnarray}
\hat{\sigma}_P^2(\tau)&=&\frac{72}{M\tau_0^2N^2(N-1)^2(N+1)^2}\sum_{i=1}^M\nonumber\\
& &\left[(D_i-D_{i+1})-\frac{N-1}{2}(C_i-C_{i+1})\right]^2\label{eq:lspvar}
\end{eqnarray}
where ($C_i$, $D_i$) and ($C_{i+1}$, $D_{i+1}$) is two pairs of accumulated sums
being consecutive. These may be either forms by the direct accumulation of
(\ref{eq:lsCsum}) and (\ref{eq:lsDsum}) or through the decimation rule of
(\ref{eq:decimateN2a}) and (\ref{eq:decimateN2b}), as long as $N$ is the
number of samples in each block (being of equal length) and that the block
observation time $\tau=N\tau_0$. Using the decimation rules, any $\tau$
calculation can be produced and then their PVAR calculated using
(\ref{eq:lspvar}). Notice that $M$ is the number of averaged blocks.

\subsection{MVAR calculation\label{sec:MVAR}}
It can be also easily verified that the sums $C_i$ may be used for computing MVAR. The MVAR estimator is defined by (\ref{eq:y_lambda}) and (\ref{eq:MVAR}). However, (\ref{eq:y_lambda}) may be rewritten as
$$
\hat{\mathbf{y}}_{k,\tau}^\Lambda=\frac{C_{k+1}-C_k}{N\tau}.
$$
Therefore, (\ref{eq:MVAR}) becomes
$$
\text{MVAR}(\tau)=\frac{1}{2N^2\tau^2}\left<\left(C_{k+2}-2C_{k+1}-C_k\right)^2\right>.
$$

\subsection{Estimator bias\label{sec:bias}}
The observant reader notices that (\ref{eq:lsfreq}) and (\ref{eq:lspvar}) does
not fully agree with the previous work. In the generic formulas $\tau=N\tau_0$
is assumed for $\tau^3$, so it would be fair to assume that $\tau^3=N^3\tau_0^3$
where this detailed analysis shows that in actual fact we should use
$\tau^3=N(N-1)(N+1)\tau_0^3=N(N^2-1)\tau_0^3$ in order to be bias-free in phase,
frequency and PVAR estimation. The scale error introduces would be $(N-1)(N+1)/N^2=1-1/N^2$, thus showing a too small value. However, due to the $N^2$ part,
this effect would be negligible for larger $N$, and this article presents a
practical way of decimating values in order to further increase $N$. However, due
warning is relevant whenever low $N$ is used, to use the corrected estimator
form. These estimators is comparable to the work of Barnes \cite{barnes1983},
but direct comparison should recall that Barnes use a shifted $t_n$ starting
with $\tau_0$ and not $0$, so due adjustments are necessary.

\section{Decimation\label{sec:soldec}}

\subsection{Decimation of different sized blocks\label{sec:dec_dif_size}}
The key idea in decimation is to form the $(C, D)$ pair for a larger set of
samples. Consider a block of $N_{12}$ samples. The definition says
\begin{eqnarray}
C_{12}&=&\displaystyle \sum_{n=0}^{N_{12}-1} x_n\\
D_{12}&=&\displaystyle \sum_{n=0}^{N_{12}-1} nx_n
\end{eqnarray}
but for practical reason processing is done on two sub blocks being $N_1$ and then $N_2$ samples long, giving
\begin{eqnarray}
N_{12}&=&N_1+N_2\\
C_1&=&\displaystyle \sum_{n=0}^{N_1-1} x_n\\
C_2&=&\displaystyle \sum_{n=0}^{N_2-1} x_{N_1+n}\\
D_1&=&\displaystyle \sum_{n=0}^{N_1-1} nx_n\\
D_2&=&\displaystyle \sum_{n=0}^{N_2-1} nx_{N_1+n}
\end{eqnarray}
the $C_{12}$ sum can be reformulated as
\begin{eqnarray}
C_{12}&=&\displaystyle \sum_{n=0}^{N_{12}-1} x_n\nonumber\\
&=&\displaystyle \sum_{n=0}^{N_1-1} x_n+\sum_{n=N_1}^{N_{12}-1} x_n\nonumber\\
&=&\displaystyle C_1+\sum_{n=0}^{N_{2}-1} x_{N_1+n}\nonumber\\
&=&\displaystyle C_1+C_2\label{eq:deciC}
\end{eqnarray}
where $N_1$ can be chosen arbitrarily under the assumption $0\le N_1\le N_{12}$ and then $N_2=N_{12}-N_1$. Similarly the $D_{12}$ sum can be reformulated as
\begin{eqnarray}
D_{12}&=&\displaystyle \sum_{n=0}^{N_{12}-1} nx_n\nonumber\\
&=&\displaystyle \sum_{n=0}^{N_1-1} nx_n+\sum_{n=N_1}^{N_{12}-1} nx_n\nonumber\\
&=&\displaystyle D_1+\sum_{n=0}^{N_{2}-1} (N_1+n)x_{N_1+n}\nonumber\\
&=&\displaystyle D_1+\sum_{n=0}^{N_{2}-1} N_1x_{N_1+n}+\sum_{n=0}^{N_{2}-1} nx_{N_1+n}\nonumber\\
&=&\displaystyle D_1+N_1\sum_{n=0}^{N_{2}-1} x_{N_1+n}+D_2\nonumber\\
&=&D_1+N_1C_2+D_2\label{eq:deciD}
\end{eqnarray}
Thus using (\ref{eq:deciC}) and (\ref{eq:deciD}) any set of consecutive blocks
can be further decimated to form a new longer block. For each decimation, only
the length $N$ and sums $C$ and $D$ needs to be stored, thus reducing the
memory requirements. In a pre\-processing stage, these sums can be produced.
The decimation rule thus allows for any length being a multiple to the pre\-processed length to be produced, with maintained non-biased phase, frequency
and PVAR estimator properties.

\subsection{Decimation by $N$}
The generalized decimate by $N$ formulation follows natural from this realization
and is proved directly though recursively use of the above rule. Consider that
a pre\-processing provides $C$ and $D$ values for block of length $N_\text{pre}$,
then on first decimation block 0 and 1 is decimated, and block 1 needs to be
raised with $N_\text{pre}C_1$ (as illustrated in Figure \ref{fig:D12}), as block 2 is decimated in the next round, $2N_\text{pre}C_2$ etc, and in general we find
\begin{eqnarray}
C_{tot}&=&\sum_{i=0}^{N_2-1}C_i\label{eq:decimateN2a}\\
D_{tot}&=&\sum_{i=0}^{N_2-1}D_i+iN_1C_i\label{eq:decimateN2b}
\end{eqnarray}
for the observation time $\tau=\tau_0N_1N_2$ with $N_1N_2$ samples, for use with the (\ref{eq:lsphase}) and (\ref{eq:lsfreq}) estimators.

This decimation by $N$ mechanism can be used together with the generic
block decimation to form any form of block processing suitable, thus providing
a high degree of freedom in how large amounts of data is being decimated.

\subsection{Geometric representation}
	\subsubsection{Decimation rule}
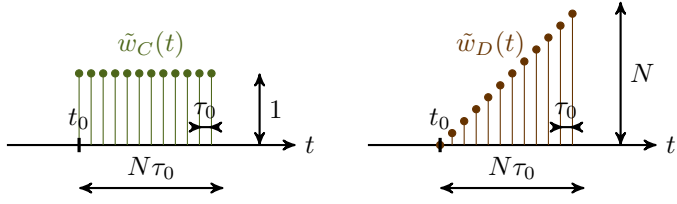
\begin{figure}[h]
\begin{tikzpicture}
\begin{scope}[scale=0.95]
\def \N {12}
\def \n {11}
\draw[thick,->,>=stealth'] (0,0)--(4,0);
\draw (4,0) node[right]{$t$};
\foreach \k in {0,...,\n}
{\draw[olive] (1+2*\k/\N,0)--++(0,1);
\draw[olive,fill] (1+2*\k/\N,1) circle(0.05);}
\draw[very thick] (1,-0.1)--(1,+.1);
\draw (1,+.1) node[above]{$t_0$};
\draw[thick,>-<,>=stealth'] (0.85+20/\N,0.25)--++(0.3+2/\N,0);
\draw (1+21/\N,0.25) node[above]{$\tau_0$};
\draw[olive] (2,1.4) node{$\tilde{w}_C(t)$};
\draw[thick,<->,>=stealth'] (1,-0.6)--(3,-0.6);
\draw (2,-0.6) node[above]{$N\tau_0$};
\draw[thick,<->,>=stealth'] (3.5,0)--(3.5,1);
\draw (3.5,0.5) node[right]{$1$};

\draw[thick,->,>=stealth'] (5,0)--(9,0);
\draw (9,0) node[right]{$t$};
\foreach \k in {0,...,\n}
{\draw[brown] (6+2*\k/\N,0)--++(0,2*\k/\N);
\draw[brown,fill] (6+2*\k/\N,2*\k/\N) circle(0.05);}
\draw[very thick] (6,-0.1)--(6,+.1);
\draw (6,+.1) node[above]{$t_0$};
\draw[thick,>-<,>=stealth'] (5.85+20/\N,0.25)--++(0.3+2/\N,0);
\draw (6+21/\N,0.25) node[above]{$\tau_0$};
\draw[brown] (6.7,1.4) node{$\tilde{w}_D(t)$};
\draw[thick,<->,>=stealth'] (6,-0.6)--(8,-0.6);
\draw (7,-0.6) node[above]{$N\tau_0$};
\draw[thick,<->,>=stealth'] (8.5,0)--(8.5,2);
\draw (8.5,1.0) node[right]{$N$};
\end{scope}
\end{tikzpicture}
\caption{weight functions of the $C$ and $D$ elementary block pair.\label{fig:CnD}}
\end{figure}

Figure \ref{fig:CnD} represents the weight functions of the $C$ and $D$ elementary block pair that we will symbolize respectively with 
{\tikz \draw[thick,olive] (0,0)--(0.2,0)--(0.2,0.15)--(0.6,0.15)--(0.6,0)--(0.8,0);}
and
{\tikz \draw[thick,brown] (0,0)--(0.2,0)--(0.6,0.3)--(0.6,0)--(0.8,0);}. 

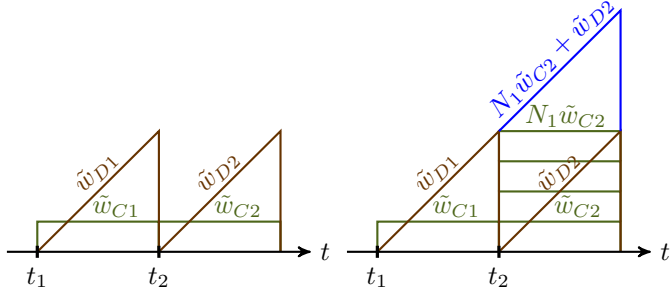
\begin{figure}[h]
\begin{tikzpicture}
\begin{scope}[scale=0.8]
\draw[thick,olive] (1,0)--(1,0.5)--(3,0.5)--(3,0)--(3,0.5)--(5,0.5)--(5,0);
\draw[olive] (2.3,0.75) node{$\tilde{w}_{C1}$};
\draw[olive] (4.3,0.75) node{$\tilde{w}_{C2}$};
\draw[thick,brown] (1,0)--(3,2)--(3,0)--(5,2)--(5,0);
\draw[brown] (2.2,1.1) node[above,rotate=45]{$\tilde{w}_{D1}$};
\draw[brown] (4.2,1.1) node[above,rotate=45]{$\tilde{w}_{D2}$};

\draw[thick,->,>=stealth'] (0.5,0)--(5.5,0);
\draw (5.5,0) node[right]{$t$};
\draw[very thick] (1,-0.1)--(1,+.1);
\draw (1,-.1) node[below]{$t_1$};
\draw[very thick] (3,-0.1)--(3,+.1);
\draw (3,-.1) node[below]{$t_2$};
\end{scope}
\end{tikzpicture}
\hfill
\begin{tikzpicture}
\begin{scope}[scale=0.8]
\draw[thick,olive] (1,0)--(1,0.5)--(3,0.5)--(3,0)--(3,0.5)--(5,0.5)--(5,0);
\draw[thick,olive] (3,0.5)--(3,1)--(5,1)--(5,0.5);
\draw[thick,olive] (3,1)--(3,1.5)--(5,1.5)--(5,1);
\draw[thick,olive] (3,1.5)--(3,2)--(5,2)--(5,1.5);
\draw[olive] (2.3,0.75) node{$\tilde{w}_{C1}$};
\draw[olive] (4.3,0.75) node{$\tilde{w}_{C2}$};
\draw[olive] (4.1,2.25) node{$N_1\tilde{w}_{C2}$};
\draw[thick,brown] (1,0)--(3,2)--(3,0);
\draw[thick,brown] (3,0)--(5,2)--(5,0);
\draw[brown] (2.2,1.1) node[above,rotate=45]{$\tilde{w}_{D1}$};
\draw[brown] (4.2,1.1) node[above,rotate=45]{$\tilde{w}_{D2}$};
\draw[thick,blue] (3,2)--(5,4)--(5,2);
\draw[blue] (4.1,3) node[above,rotate=45]{$N_1\tilde{w}_{C2}+\tilde{w}_{D2}$};

\draw[thick,->,>=stealth'] (0.5,0)--(5.5,0);
\draw (5.5,0) node[right]{$t$};
\draw[very thick] (1,-0.1)--(1,+.1);
\draw (1,-.1) node[below]{$t_1$};
\draw[very thick] (3,-0.1)--(3,+.1);
\draw (3,-.1) node[below]{$t_2$};
\end{scope}
\end{tikzpicture}

\caption{Decimation rule of the $(C_1,D_1)$ and $(C_2,D_2)$ block pair weight functions over two adjacent sub-sequences for composing the $(C_{12},D_{12})$ block pair weight functions over the whole sequence.\label{fig:D12}}
\end{figure}

In the same way as in {\S} \ref{sec:dec_dif_size}, let us consider two consecutive sets of $N_1$ samples, beginning respectively at instants $t_1$ and $t_2$, and the whole sequence of $N_{12}=2N_1$ samples. We can form the blocks $C_1$ and $D_1$ over the first sub-sequence, $C_2$ and $D_2$ over the second one as well as $C_{12}$ and $D_{12}$ over the whole sequence (see left hand side of Figure \ref{fig:D12}). The right hand side of Figure \ref{fig:D12} shows that $C_{12}=C_1+C_2$ and $D_{12}=D_1+N_1C_2+D_2$ as demonstrated in (\ref{eq:deciC}) and (\ref{eq:deciD}).

	\subsubsection{The $\Omega$-counter weight function}

\begin{figure}[h]
\begin{center}
\begin{tikzpicture}
\begin{scope}[scale=1]
\draw[thick,olive] (1,0)--(1,-0.5)--(5,-0.5)--(5,0);
\draw[thick,olive] (1,-0.5)--(1,-1)--(5,-1)--(5,-0.5);
\draw[olive] (3,-1.12) node[above]{$-N\tilde{w}_C/2$};
\draw[thick,brown] (1,0)--(5,2)--(5,0);
\draw[brown] (3,.95) node[above,rotate=26.565]{$\tilde{w}_D$};
\draw[thick,blue] (1,0)--(1,-1)--(5,1)--(5,0);
\draw[blue] (4.15,0.455) node[above,rotate=26.565]{$\tilde{w}_D-N\tilde{w}_C/2$};

\draw[thick,->,>=stealth'] (0.5,0)--(5.5,0);
\draw (5.5,0) node[right]{$t$};
\draw[very thick] (1,-0.1)--(1,+.1);
\draw (1,.1) node[above]{$-\tau/2$};
\draw[very thick] (3,-0.1)--(3,+.1);
\draw (3,.1) node[above]{$0$};
\draw[very thick] (5,-0.1)--(5,+.1);
\draw (5,.1) node[above]{$+\tau/2$};
\end{scope}
\end{tikzpicture}
\caption{Association of the $(C,D)$ block pair weight functions for composing the $\Omega$-counter weight function.\label{fig:Om_wf}}
\end{center}
\end{figure}
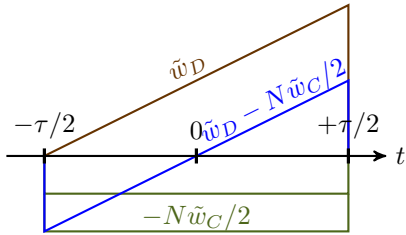

As stated in {\S} \ref{sec:Omega_and_PVAR}, the weight function of the $\Omega$-counter for phase data $x(t)$ is given by (\ref{eq:wcx}). Figure \ref{fig:Om_wf} shows that the estimate $\hat{\mathbf{y}}^\Omega$ of the $\Omega$-counter is:
$$
\hat{\mathbf{y}}^\Omega\propto D - \frac{N}{2} C.
$$

	\subsubsection{The PVAR weight function}
\begin{figure}[!h]
\begin{center}
\begin{tikzpicture}
\begin{scope}[scale=0.95]
\draw[thick,olive] (1,0)--(1,0.5)--(5,0.5)--(5,0);
\draw[thick,olive] (1,0.5)--(1,1)--(5,1)--(5,0.5);
\draw[olive] (2,0.885) node[above]{$+N\tilde{w}_{C1}/2$};
\draw[thick,brown] (1,0)--(5,2)--(5,0);
\draw[brown] (4,1.43) node[above,rotate=26.565]{$\tilde{w}_{D1}$};
\draw[thick,blue] (1,0)--(1,1)--(5,-1)--(5,0);
\draw[blue] (3.5,-0.37) node[above,rotate=-26.565]{$N\tilde{w}_{C1}/2-\tilde{w}_{D1}$};

\draw[thick,olive] (5,0)--(5,-0.5)--(9,-0.5)--(9,0);
\draw[thick,olive] (5,-0.5)--(5,-1)--(9,-1)--(9,-0.5);
\draw[olive] (7,-1.1) node[above]{$-N\tilde{w}_{C2}/2$};
\draw[thick,brown] (5,0)--(9,2)--(9,0);
\draw[brown] (8,1.43) node[above,rotate=26.565]{$\tilde{w}_{D2}$};
\draw[thick,blue] (5,0)--(5,-1)--(9,1)--(9,0);
\draw[blue] (7.95,0.36) node[above,rotate=26.565]{$\tilde{w}_{D2}-N\tilde{w}_{C2}/2$};

\draw[thick,->,>=stealth'] (0.5,0)--(9.5,0);
\draw (9.5,0) node[right]{$t$};
\draw[very thick] (1,-0.1)--(1,+.1);
\draw (1,-.1) node[below]{$-\tau$};
\draw[very thick] (5,-0.1)--(5,+.1);
\draw (5.1,-.1) node[below]{$0$};
\draw[very thick] (9,-0.1)--(9,+.1);
\draw (9,-.1) node[below]{$+\tau$};
\end{scope}
\end{tikzpicture}
\caption{Association of the $(C_1,D_1)$ and $(C_2,D_2)$ block pair weight functions for composing the PVAR weight function.\label{fig:PVAR_wf}}
\end{center}
\end{figure}
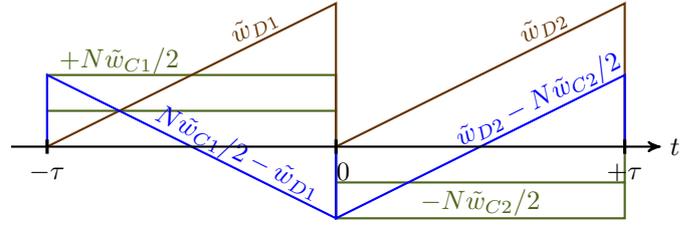

Since PVAR$(\tau)=\frac{1}{2}\left<\left(\hat{\mathbf{y}}^{\Omega}_{2}-\hat{\mathbf{y}}^{\Omega}_{1}\right)^2\right>$, it comes
$$
\text{PVAR}(\tau)\propto\left<\left(D_2-\frac{N}{2}C_2-D_1+\frac{N}{2}C_1\right)^2\right>
$$
as illustrated by Figure \ref{fig:PVAR_wf}.

\subsection{Decimation processing}
It should be noted that the decimation process may be used
recursively, such that it is used as high-speed pre\-processing in FPGA and that
the $(C, D)$ pairs is produced for each $N_1$ samples as suitable for the
plotted lowest $\tau$. Another benefit of the decimation processing is that if
the FPGA front-end has a limit to the number of supported $N_1$ it can process,
software can then continue the decimation without causing a bias. This provides
for a high degree of flexibility without suffering from high memory
requirements, high processing needs or for that matter overly complex HW
support.

\subsection{Decimation biases}
%

The decimation process avoids the low-$\tau$ bias errors typically seen when
using frequency estimation with $\Lambda$-counters or $\Omega$-counters. Such
setup suffers from the fact that the counter have a fixed $\tau=1/(2\pi f_H)$ pre-filter of
$\Lambda$ or $\Omega$ type, which reduces the system bandwidth $f_H$, but for
higher $\tau$ the main lobes of AVAR frequency response is within the
pre-filter pass-band of this system bandwidth and the pre-filtering provides no
benefit for longer $\tau$ readings, for such $\tau$ values the raw phase or
frequency readings of the $\Lambda$-counter mode can be used directly as the
fixed $\tau$ $\Lambda$ or $\Omega$ pre-filtering adds no benefit.

The pre-filtering is indeed to filter out white noise for better frequency
readings, but the AVAR processing is unable to extend the filtering to higher
$\tau$ values, as it does not perform $\tau$-filtering decimation. The proposed
decimation rule is able to provide proper $\Omega$ response for any $\tau$, but
bias-free decimation is only possible using two scalar values rather than one.

Thus, existing $\Omega$-counters producing a single frequency reading for a
block interval is not possible to decimate properly. Existing counters is best
used in their $\Pi$ time-interval mode and decimation performed in software
instead. The scalar values $C$ and $D$ can also be used to estimate phase and
frequency with least square properties, so this comes as a benefit for normal
counter usage.

Any users of $\Lambda$-counter and $\Omega$-counter should be cautioned to use
them without proper decimation routines, as their estimates can be biased and
unusable for metrology use.

\subsection{Multi-$\tau$ decimation}
Another aspect of the decimation processing is not only that many $\tau$ can be
produced out of the same sample or block sequence, but once a suitable set of
$\tau$ variants have been produced, these can be decimated recursively in
suitable form to create $\tau$ variants of higher multiples. One such approach
would be to produce the 1 to 9 $\tau$ multiples (or only 1, 2 and 5 multiples which is enough for a log-log plot) of accumulates for $\tau$ being
1 s, thus producing the 1 to 9 s sums, and by recursive decimation by 10
produces the same set of points on the log-log plot, but for 10$^\text{th}$ multiple of
time, for each recursive step. This will allow for large ranges of $\tau$ to be
calculated for a reasonable amount of memory and calculation power.

\subsection{Overlapping decimation}
It should be realized that decimation over say 10 blocks can produce new
decimation values for each new block that arrives, thus providing an
overlapping process. Doing such overlapping ensures that the achieved
degrees of freedom remains high and thus it is the recommended process for post\-processing.

On the other hand, the necessity to achieve the highest sampling rate prohibits the use of overlapping for pre\-processing. The resulting loss on degrees of freedom is advantageously compensated by the benefits of high speed accumulation which ensures that white phase noise is being reduced
optimally in order to improve the quality of higher $\tau$ data.

Users shall be cautioned not to interleave the intermediate blocks, as this
paper does not detail how decimation should be performed with overlapped
results. The decimation routines assumed continuous blocks, but not overlapping
blocks. The improper use of overlapping blocks will produce biases in estimates
and should be avoided. If overlapping, and thus multiple reference to phase
samples, is avoided even considering interleaved processing, the biasing
effect can be avoided.

\subsection{Hardware implementation issues}

\subsubsection{$\tau_0$}
A hardware/FPGA implementation will time-stamp every $N_\text{prescale}$ cycle of the
signal. The period $t_\text{period}=1/f$ of the incoming signal together with the
prescale division provide the basic observation interval
$\tau_0=N_\text{prescale}\times t_\text{period}$. Keeping the $\tau_0$ low ensures that the
white noise rejection and the counter quantization noise rejection, both
following the $1/{\tau^{1.5}}$ deviation slope, gets into action quickly and
rejects these noises such that actual source noise can be observed for short
tau.

\subsubsection{Noise rejection}
Consider a 10 MHz source, where we time-stamp every period in a 100 MHz
clock with no interpolation. The quantization noise can be estimated to be
$1/(\sqrt{12}\tau^{1.5})$, as illustrated in Table \ref{tab:noisereject},
\begin{table}
\caption{Illustration of counter noise reduction with $\tau$.\label{tab:noisereject}}
\begin{center}
\begin{tabular}{|r|r|}      \hline
\multicolumn{1}{|c|}{$\tau$} & \multicolumn{1}{c|}{noise} \\ \hline
$100$ ns     & $1.6E-9$  \\ \hline
  $1$ $\mu$s  & $5E-11$   \\ \hline
 $10$ $\mu$s  & $1.6E-12$ \\ \hline
$100$ $\mu$s  & $5E-14$   \\ \hline
  $1$ ms     & $1.6E-15$ \\ \hline
\end{tabular}
\end{center}
\end{table}
thus providing a high rejection of noise already at 1 ms
observation intervals, even if no hardware interpolation is done. The
interpolation is instead done using a very high amount of samples (10 MS/s in
this scenario) which is least-square matched, thus rejecting the quantization
noise. The white noise rejection follows the same properties. It should be
understood that the decimation will step-wise provide a narrower system
bandwidth, as expected from MVAR and PVAR processing.

\subsubsection{Time-stamping}
The time-stamping for each event, every $\tau_0$, is done by sampling a
free-running time-counter, that forming the time sample of that event.
As we decimate these samples into a block, we run into the aspect of the
wrapping of the time counter. If the time-counter is large enough, it may wrap
once within a block. By keeping the first time-stamp, then if adding $N-1$
causes it to go beyond the wrap-count, then we know that the counter wrapped.

\section{Towards a standard for the output format of $\Omega$-counters\label{sec:standard}}
	\subsection{Basic $\Omega$-counter}
In order to maximize the acquisition speed of such a counter, the pre\-processing should be reduced to its minimum. Therefore, we recommend to only compute and store the $(C,D)$ block pair at each initial step without any normalization. The length of the step $\tau_0$ should be chosen in order to have a knowledge of what happens at short term without the storage rate of the block pairs becomes an issue. A duration range between $1$ ms $<\tau_0< 100$ ms seems to be good compromise.

In order to reach the best phase noise rejection, the smallest $t_\text{period}=1/f$ should be chosen. Thus, for a DUT frequency of $10$ MHz, this yield:
\begin{itemize}
	\item $t_\text{period}=100$ ns
	\item $\tau_0=10$ ms
	\item $N_\text{prescale}=10^5$.
\end{itemize}

	\subsection{An universal counter}
We noticed in {\S} \ref{sec:MVAR} that MVAR can be computed from the $C_i$ blocks. Therefore, an $\Omega$-counter provides also the basic data for calculating MVAR and thus may be considered as an improved $\Lambda$-counter. Moreover, by adding the initial phase of each pre\-processing block, i.e. $x_i$, it will be also possible to compute directly AVAR. Such a counter, providing at each step the triplet $(x_i, C_i, D_i)$, may be considered as an ``universal counter''.

\section{Summary}

Presented is an improved method to perform least-square phase, frequency and
PVAR estimates, allowing for high speed accumulation similar to \cite{snyder1981}, but extending into any $\tau$ needed. It also provides for multi-$\tau$
analysis from the same basic accumulation. The decimation method can be applied
recursively to form longer $\tau$ estimates, reusing existing calculations and
thus saving processing. Thus, it provides a practical method to provide PDEV
log-log plots, providing means to save memory and processing power without the
risk of introducing biases in estimates, as previous methods have shown.

\bibliography{pddec.bib}

\end{document}